\documentclass[a4paper,12pt]{article}

\usepackage{amsmath}
\usepackage[final]{graphicx}
\usepackage{bm}
\usepackage{amssymb}
\usepackage{color}
\usepackage{amssymb}
\newcounter{comment}

{\refstepcounter{comment}%
\begin{quote}
\ttfamily\small$\blacksquare$ \textbf{\underline{Comment} $\sharp$\thecomment:}}%
{\end{quote}}

\begin{document}
\hfill
\begin{minipage}{20ex}\small
ZTF-EP-14-03\\
\end{minipage}

\begin{center}
{\LARGE \bf Electroweak breaking and Dark Matter from the
common scale\\}
\vspace{0.7in}

{\bf Sanjin~Beni\'c and Branimir~Radov\v{c}i\'c\\}
\vspace{0.1in}
{\sl Department of Physics, Faculty of Science, University of Zagreb,\\
P.O.B. 331, HR-10002 Zagreb, Croatia\\}
\vspace{0.7in}
\today \\[5ex]
\end{center}

\vspace{0.2in}

\begin{abstract}

We propose a classically scale invariant extension of the Standard
Model where the electroweak symmetry breaking and the mass of the
Dark Matter particle come from the common scale. We introduce
$U(1)_X$ gauge symmetry and $X$-charged scalar $\Phi$ and Majorana fermion
$N$. Scale invariance is broken via Coleman-Weinberg mechanism pro-
viding the vacuum expectation value of the scalar $\Phi$.
Stability of the dark
matter candidate $N$ is guaranteed by a remnant $Z_2$ symmetry. The
Higgs boson mass and the mass of the Dark Matter particle have a
common origin, the vacuum expectation value of $\Phi$.
Dark matter relic
abundance is determined by annihilation $NN \to \Phi\Phi$. We scan the
parameter space of the model and find the mass of the dark matter
particle in the range from 500 GeV to a few TeV.

\end{abstract}

\clearpage

\section{Introduction}

The discovery of the Higgs boson \cite{Aad:2012tfa,Chatrchyan:2012ufa} is a dramatic confirmation of the Stan-
dard Model (SM). Despite finding last missing piece of SM we are still looking
for insight into the detailed mechanism of electroweak symmetry breaking.
In particular, within SM, the Higgs mass receives large corrections leading
to the hierarchy problem. For several resolutions of the hierarchy problem
that have been put forward, like supersymmetry, large extra dimensions or
composite structure the LHC found no hints so far.

Given such a scenario where new physics, stemming from e. g. supersym-
metry, seems to be absent, one is lead to pursue alternative approaches to the
hierarchy problem. Here, we follow an idea first put forward by Bardeen \cite{Bardeen:1995kv}
that in classically scale invariant (SI) theories, scale invariance is broken by
quantum corrections and quadratic divergence in fundamental scalar masses
is a cutoff regularization artefact. All dimensionfull parameters in these the-
ories, including the scalar masses, come from a single renormalization scale.
As has been shown by Coleman and E.~Weinberg (CW) \cite{Coleman:1973jx} in order for this
mechanism of mass generation to work within perturbation theory at least
two bosonic degrees of freedom are needed. A seminal paper by Gildener and
S.~Weinberg \cite{Gildener:1976ih} (GW) represents a first cohesive study of the CW mechanism
for multiple scalars.

Classical scale invariance breaks by quantum anomaly yielding one pseudo-
Goldstone boson, the so-called scalon \cite{Gildener:1976ih}. The intriguing idea of identifying the
scalon with the Higgs particle requires additional bosonic degrees
of freedom due to the large top quark mass. Sprung by initial explorations \cite{Meissner:2006zh}
many works have been put forward to realize this scenario. In order to
compensate the large top contribution, extra bosons either have large couplings
to the Higgs \cite{Foot:2007as,AlexanderNunneley:2010nw,Antipin:2013exa} or appear in large multiplets \cite{Espinosa:2007qk}.
Alternatively, the CW
mechanism can be applied in a new gauge sector where the scale
gets transmitted to the SM through the Higgs portal \cite{Hempfling:1996ht,Chang:2007ki,Iso:2009ss,Englert:2013gz,Hambye:2013dgv,Carone:2013wla}. For additional
work on the SI extensions of SM in different contexts, see
\cite{Foot:2007ay,Foot:2007iy,Meissner:2008gj,Holthausen:2009uc,Dermisek:2013pta}.

Dark matter (DM) is another problem pressing us to extend the SM. Classically
SI extensions of the SM with DM have been studied in inert Higgs doublet
model \cite{Hambye:2007vf}, $SU(2)$ vector dark matter \cite{Hambye:2013dgv,Carone:2013wla}, multiple scalar models \cite{AlexanderNunneley:2010nw,Foot:2010av,Ishiwata:2011aa,Farzinnia:2013pga,Gabrielli:2013hma,Guo:2014bha} and
non-perturbative realizations of the hidden sector \cite{Hur:2011sv,Heikinheimo:2013fta,Holthausen:2013ota}.

SI theories and the CW mechanism offer a possibility of a single origin
of the electroweak symmetry breaking and the dark matter mass embodying
the paradigm of the WIMP miracle. This attractive picture is the subject
of our work. We include SI dark sector consisting of a new $U(1)_X$ gauge
group and $X$-charged scalar $\Phi$ and Majorana fermion $N$. The effective scalar
potential is dominated by the contributions of the new gauge boson and leads
to the breaking of the dark gauge symmetry. Via the Higgs portal coupling the
CW mechanism induces the breaking of electroweak symmetry. Due to the
residual $Z_2$ symmetry the Majorana fermion is stable and represents the DM
candidate.

\section{The model}

We introduce $U(1)_X$ gauge symmetry with doubly $X$-charged scalar $\Phi$ and
singly $X$-charged Majorana fermion $N$, both singlets under the SM gauge
group. The scalar potential and dark Yukawa term are
\begin{equation}
V(H,\Phi) = \frac{\lambda_H}{2}(H^\dag H)^2+\frac{\lambda_\Phi}{2}(\Phi^\dag\Phi)^2+
\lambda_P(H^\dag H) (\Phi^\dag\Phi)~,
\label{eq:pot}
\end{equation}
\begin{equation}
\mathcal{L}_y = -\frac{y}{2}\Phi\bar{N}N~,
\end{equation}
where $H$ is a SM Higgs doublet.

The CW mechanism will be studied in the GW framework \cite{Gildener:1976ih} where quantum corrections are built on top of
a flat direction in the tree-level potential
giving mass to the scalon. Assuming that both $H$ and $\Phi$ acquire vacuum
expectation values (\emph{vevs})
\begin{equation}
H
=
\begin{pmatrix}
H^+\\
\frac{1}{\sqrt{2}}(v_H+h'+iG)\\
\end{pmatrix}
\, , \qquad \Phi = \frac{1}{\sqrt{2}}(v_\Phi+\phi'+iJ)~,
\end{equation}
the potential (\ref{eq:pot}) has a flat direction when the couplings
satisfy
\begin{equation}
\lambda_H(\Lambda)\lambda_\Phi(\Lambda)-\lambda_P^2(\Lambda)=0~.
\end{equation}
This is a dimensional transmutation, where one dimensionless
coupling is traded
for the physical scale of the model $\Lambda$.

At the tree level the \emph{vevs} are given by
\begin{equation}
\frac{v_H^2}{v_\Phi^2} = - \frac{\lambda_P}{\lambda_H}~.
\label{eq:type2}
\end{equation}
Scalar mass eigenstates are given by
\begin{equation}
\begin{pmatrix}
h\\
\phi\\
\end{pmatrix}
=
\begin{pmatrix}
\cos\theta & -\sin\theta\\
\sin\theta & \cos\theta\\
\end{pmatrix}
\begin{pmatrix}
h'\\
\phi'\\
\end{pmatrix}~,
\end{equation}
with mixing angle $\theta$ given by
\begin{equation}
\sin^2\theta = -\frac{\lambda_P}{\lambda_H-\lambda_P}~.
\label{eq:cos}
\end{equation}
The tree-level masses of scalars are
\begin{equation}
m_h^2 = (\lambda_H - \lambda_P) v_H^2 \, , \qquad m_\phi^2 = 0~,
\end{equation}
and $X$-gauge boson and fermion masses are
\begin{equation}
m_X^2 = 4g_X^2 v_\Phi^2 \, , \qquad m_N = \frac{y}{\sqrt{2}}v_\Phi
\end{equation}
where $g_X$ is the gauge coupling of the $U(1)_X$ gauge symmetry.

To find the scalon mass we need the one-loop corrected potential along the flat direction which takes
the following form \cite{Gildener:1976ih}
\begin{equation}
\delta V(r) = A r^4 + B r^4 \log\left(\frac{r^2}{\Lambda^2}\right)~,
\label{eq:one_loop}
\end{equation}
where the radial field $r$ is defined through
\begin{equation}
\begin{pmatrix}
v_H+h'\\
v_\Phi+\phi'\\
\end{pmatrix}
=r
\begin{pmatrix}
n_h\\
n_\phi\\
\end{pmatrix}
~,
\end{equation}
while the fields $n_h,n_\phi$ satisfy
the constraint $n_h^2 + n_\phi^2 = 1$, with their \emph{vevs} being
$\sin\theta$ and $\cos\theta$, respectively.
The coefficients $A$ and $B$ are
given as \cite{AlexanderNunneley:2010nw}
\begin{equation}
\begin{split}
A&=\frac{1}{64\pi^2 v_r^4}\Bigl\{m_h^4 \left(-\frac{3}{2}+\log\frac{m_h^2}{v_r^2}\right)+6m_W^4 \left(-\frac{5}{6}+\log\frac{m_W^2}{v_r^2}\right)\\
&+3m_Z^4 \left(-\frac{5}{6}+\log\frac{m_Z^2}{v_r^2}\right)+3m_X^4 \left(-\frac{5}{6}+\log\frac{m_X^2}{v_r^2}\right)\\
&-12m_t^4 \left(-1+\log\frac{m_t^2}{v_r^2}\right)-2m_N^4 \left(-1+\log\frac{m_N^2}{v_r^2}\right)\Bigr\}~,
\end{split}
\end{equation}
\begin{equation}
B = \frac{1}{64\pi^2 v_r^4}\left(m_H^4+6m_W^4+3m_Z^4+3m_X^4-12m_t^4-2m_N^4\right)~,
\end{equation}
where $v_r$ is the \emph{vev} of the field $r$.
The mass of the scalon is given as
\begin{equation}
m_\phi^2 = \frac{\partial^2 \delta V}{\partial r^2}\Big|_{r=v_r}
=8Bv_r^2~.
\label{eq:scalmass}
\end{equation}

The tree-level potential (\ref{eq:pot}) in the mass eigenstate basis reads
\begin{equation}
\begin{split}
V(h,\phi) &= \frac{1}{2}m_h^2 h^2 + \frac{1}{2}\sqrt{1-\frac{\lambda_P}{\lambda_H}}(\lambda_P+\lambda_H)v_H h^3 + \frac{1}{8}\frac{(\lambda_H+\lambda_P)^2}{\lambda_P}h^4\\
&+\sqrt{-\lambda_P(\lambda_H-\lambda_P)}v_H h^2 \phi +\frac{1}{2}\sqrt{\frac{-\lambda_P}{\lambda_H}}(\lambda_H+\lambda_P)h^3 \phi-\frac{1}{2}\lambda_P h^2\phi^2~.
\end{split}
\end{equation}
Notice that the couplings $h\phi^2$, $\phi^3$, $h\phi^3$
and $\phi^4$ vanish at the tree level \cite{Chang:2007ki,Farzinnia:2013pga},
independent of the particular value of the Higgs portal
coupling $\lambda_P$.
In particular,
the rigorous GW treatment for multiple scalar
fields states that in the model
studied here there are no Higgs decays to the scalons
on the tree level \cite{Chang:2007ki,Farzinnia:2013pga},
contrary to \cite{Englert:2013gz}.
These points can be understood from the fact that flat direction of the tree-level potential defined by the \emph{vev} of
the unit vector field $(n_h,n_\phi)$ is also an eigenvector of the
mass matrix with zero eigenvalue.
Therefore, if a term in the tree-level potential has more than two
physical $\phi$ fields, or two $\phi$ fields
and a dimensionfull coupling, the respective coefficient will be zero
by construction.

Due to the remnant $Z_2$ symmetry the Majorana fermion $N$ is a DM
candidate \cite{Ma:2013yga}. Our DM scenario assumes that $N$ is heavier than the scalar $\phi$.
In the early Universe DM annihilates dominantly through the $t$-channel to $\phi\phi$ pair,
while contributions suppressed by a small mixing arise from $NN \to hh$ and
$NN \to h\phi$ processes. At the time of the decoupling of the DM, scalar $\phi$ is in
the thermal equilibrium with the thermal bath of the SM particles through
the Higgs portal interaction. While our numerical analysis covers all the
mentioned annihilation processes, it is instructive to show the dominant one.
The thermally averaged $p$-wave annihilation cross section for $NN \to \phi\phi$ is
\begin{equation}
\langle \sigma (NN \to \phi\phi) v \rangle = \frac{y^4 \cos^4\theta}{96\pi}
\sqrt{1-\frac{m_\phi^2}{m_N^2}} \frac{m_N^2(9m_N^4-8m_N^2 m_\phi^2 + 2m_\phi^4)}
{(2m_N^2-m_\phi^2)^4}v^2~.
\label{eq:dmcs}
\end{equation}
In our calculations we use the
observed value for the DM relic density of the Universe,
$\Omega_{DM} h^2 = 0.1187(17)$~\cite{Ade:2013zuv}.

\section{Results}

The model introduces four new parameters: gauge coupling $g_X$, Yukawa coupling
$y$ and quartic scalar couplings $\lambda_P$ and $\lambda_\Phi$.
The two constraints given
by the Higgs boson mass and the dark matter relic abundance leave two
undetermined parameters chosen to be $m_X$ and $m_\phi$.
The mass of the $X$-boson has to be larger than $\sim 600$ GeV to overcome the
top quark contribution in order to have positive mass for the scalon. Higgs
searches at LEP exclude scalon masses under 114 GeV in our model. Global
fit to the current LHC data \cite{Farzinnia:2013pga} gives an upper
bound of $\sin\theta < 0.37$.
We scan the parameter space of the model from the initial values $m_X = 600$
GeV, $m_\phi = 114$ GeV, up to $m_X = 2000$ GeV, $m_\phi = 400$ GeV.

\begin{figure}[h]
\centering
\centerline{\includegraphics[scale=0.53]{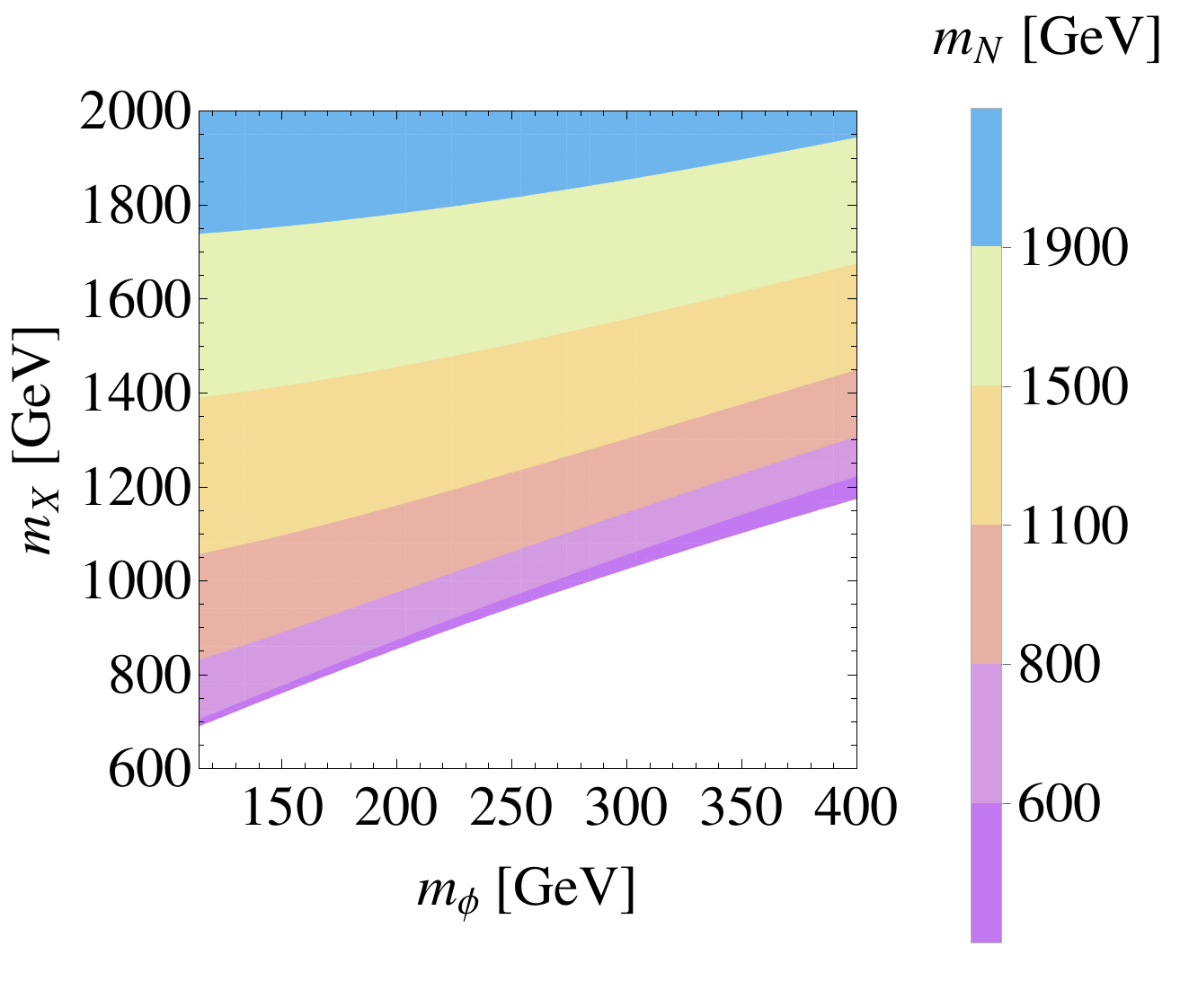}
\hspace{0.25cm}
\includegraphics[scale=0.53]{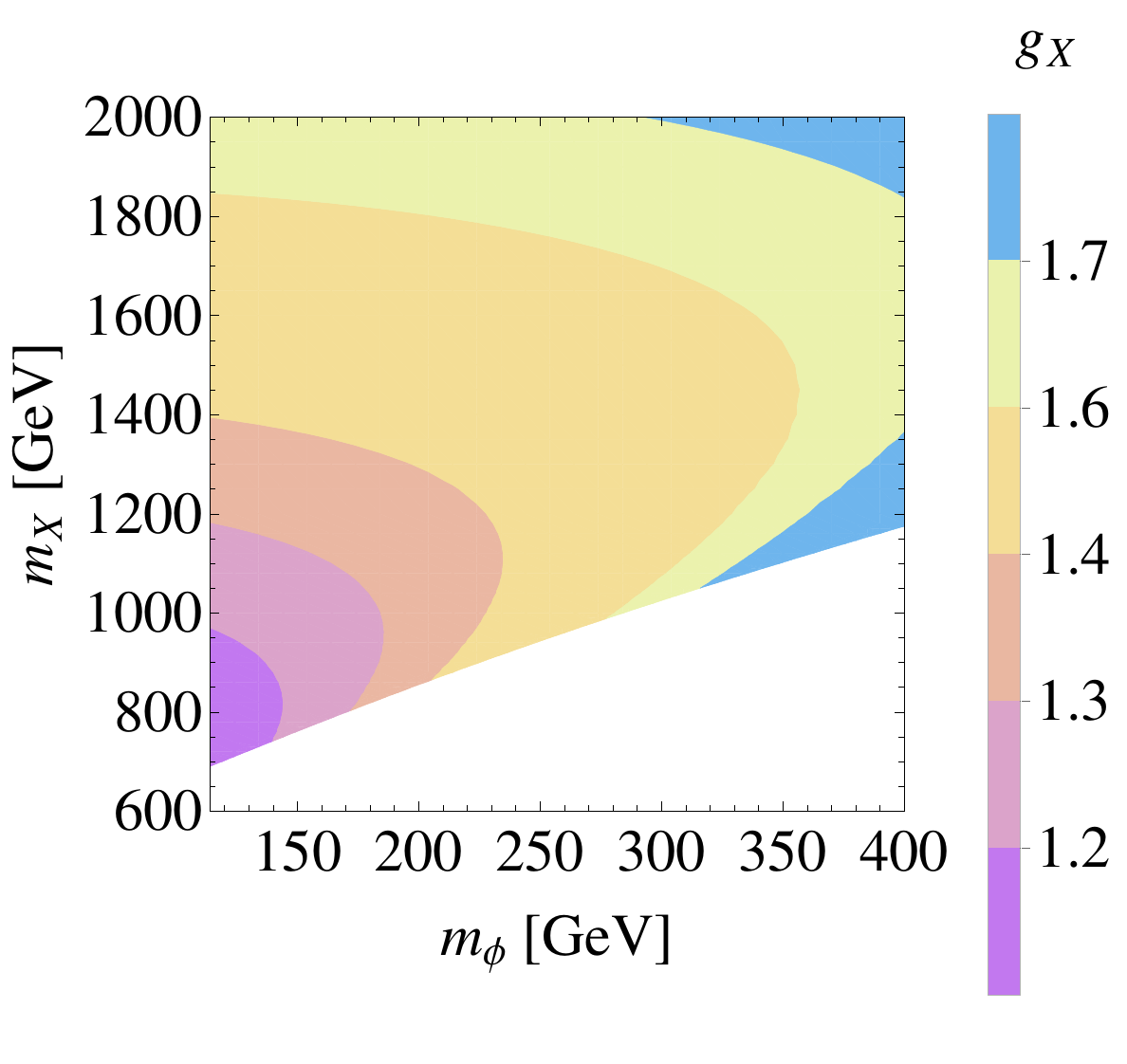}}
\caption{Mass of the DM candidate $m_N$ (left) and dark sector gauge coupling
$g_X$ (right) as a function of dark gauge boson mass $m_X$ and the scalon mass
$m_\phi$. The lower region on the plot is excluded by the LHC
bound on $\sin\theta$.}
\label{fig:gm}
\end{figure}

\begin{figure}[h]
\centering
\centerline{\includegraphics[scale=0.53]{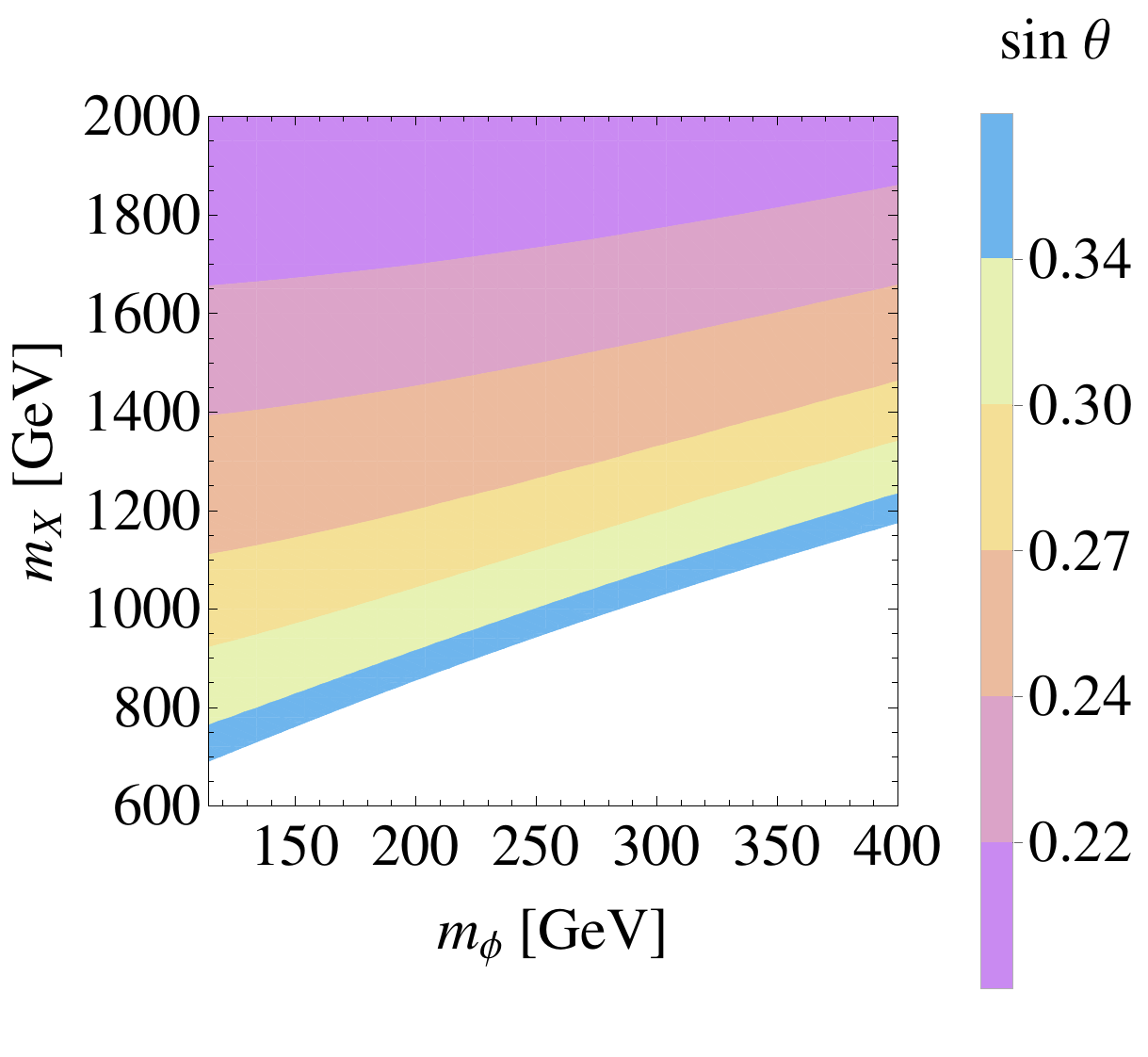}
\hspace{0.25cm}
\includegraphics[scale=0.53]{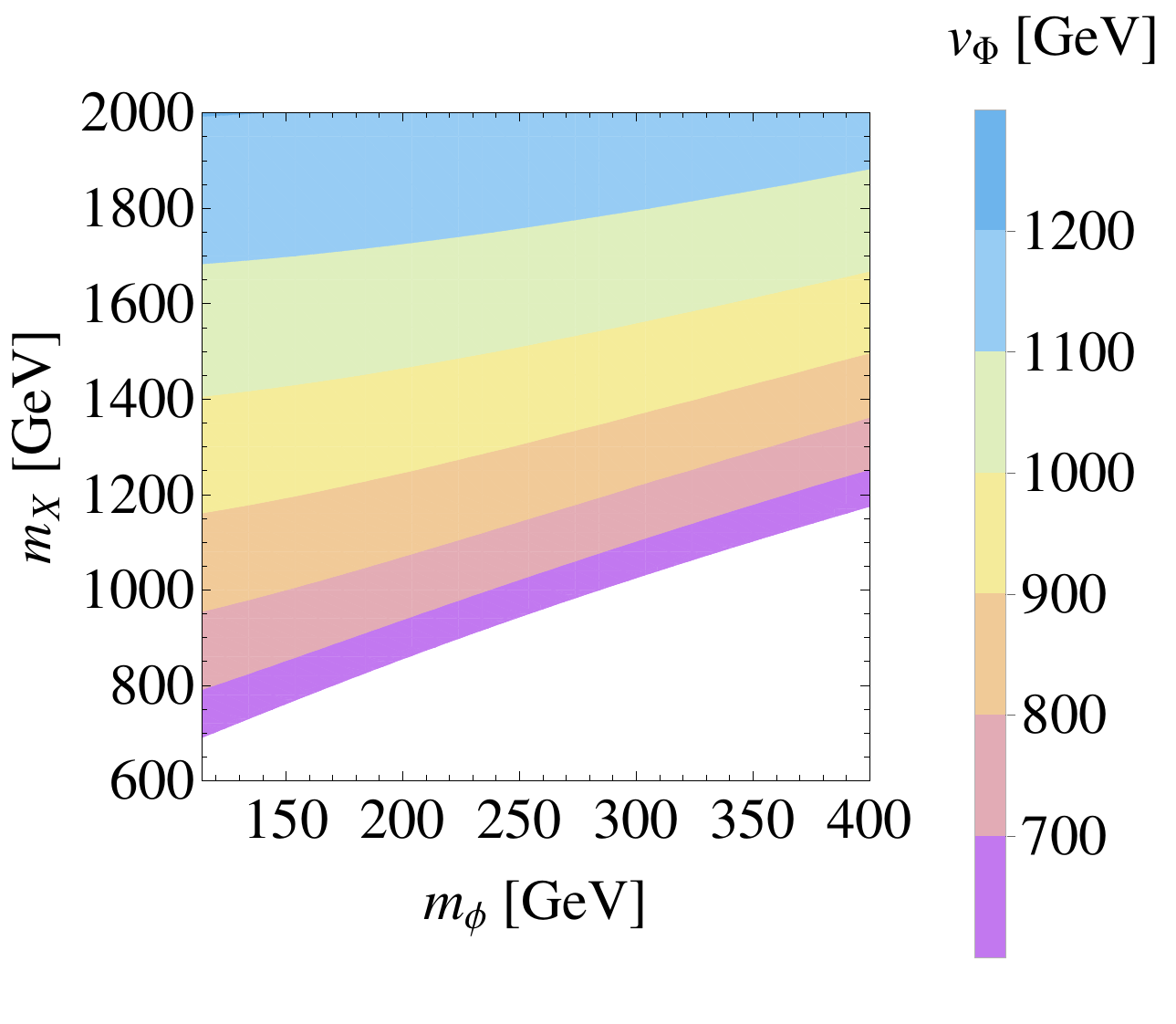}}
\caption{Mixing angle $\sin\theta$ (left) and the \emph{vev} of $\Phi$, $v_\Phi$ (right) as a function
of dark gauge boson mass $m_X$ and the scalon mass $m_\phi$. The lower region on the plot is excluded by the LHC
bound on $\sin\theta$.}
\label{fig:sin_vphi}
\end{figure}

We give our results as prediction for various quantities in the $m_X - m_\phi$
plane. On Fig.~\ref{fig:gm} we give mass of the DM candidate $m_N$ and dark sector
gauge coupling $g_X$. On Fig.~\ref{fig:sin_vphi} we
give the mixing angle $\sin\theta$
and vev of $\Phi$,
$v_\Phi$. Holding $m_\phi$ fixed while increasing $m_X$ increases the \emph{vev}
$v_\phi$ naturally leading to a smaller mixing angle $\theta$.

From (\ref{eq:dmcs}) we see that the scale $v_\Phi$
dominates the DM annihilation cross section
\begin{equation}
\langle\sigma v\rangle\sim y^4/m_N^2\sim y^2/v_\Phi^2 \ .
\end{equation}
Therefore, in order to keep $\langle\sigma v\rangle$
fixed to its value provided by the DM relic
abundance, $y$ increases when $v_\Phi$ increases.
The CW mechanism then requires
that $g_X$ is increased to have a positive scalon mass. Notice the key role played
by the DM constraint; keeping $m_\phi$ fixed and increasing $m_X$ leads to an
increase of $g_X$.
This is in contrast to the model without DM where one naively expects
$m_\phi \sim g_X m_X$.
In other words, in the regime of large $v_\Phi$ the
Majorana fermion $N$ starts to play an important role in the CW mechanism.

For the minimal value $v_\Phi\sim 700$ GeV, the
DM constraint requires a relatively large Yukawa coupling.
The CW mechanism pulls $g_X$ in the same direction,
which is the origin of $g_X$ being of order one.
The perturbative restrictions provide upper bounds on
masses in the hidden sector.
We find the mass of the DM to be in the range from $500$ GeV up to a few TeV.

\section{Conclusion}

Postulating a hidden $U(1)_X$ sector consisting
of a doubly
$X$-charged scalar and a singly charged Majorana fermion
we have examined
a SI extension of the SM with Majorana fermion
as a DM candidate.
Using the GW approach we have calculated masses of the new
particles.
The Majorana fermion saturates the DM relic abundance
via the $NN\to\phi\phi$ annihilation.

We predict the DM particle mass to be in the range from 500 GeV
to a few TeV.
While the lower bound is obtained by the LHC limit on $\sin\theta$, the upper bound is a conservative estimate set by moderate values of the dark gauge and Yukawa
couplings.
The DM constraint plays an essential role here as it requires
an appreciable Yukawa coupling.
In this sense similar results are
obtained by \cite{Hambye:2013dgv,Carone:2013wla}
where DM constraint
enforces a gauge coupling of order one.

A general statement may be drawn for a SI extension of SM
with a new gauge group;
the couplings in the hidden
sector can be taken to be small provided
the \emph{vev} of the hidden scalar is large enough.
However, identifying possible stable states of the hidden sector with
DM particles, the
scale set by CW mechanism requires moderate couplings in the hidden
sector.

\subsubsection*{Acknowledgments}
This work is supported by the
University of Zagreb under Contract No.~202348 and by the Croatian Ministry of Science, Education and
Sports under Contract No. 119-0982930-1016.


\begin{thebibliography}{10}

\bibitem{Aad:2012tfa}
  G.~Aad {\it et al.}  [ATLAS Collaboration],
  \emph{Observation of a new particle in the search for the Standard Model Higgs boson with the ATLAS detector at the LHC},
  Phys.\ Lett.\ B {\bf 716} (2012) 1
  [arXiv:1207.7214 [hep-ex]].

\bibitem{Chatrchyan:2012ufa}
  S.~Chatrchyan {\it et al.}  [CMS Collaboration],
  \emph{Observation of a new boson at a mass of 125 GeV with the CMS experiment at the LHC},
  Phys.\ Lett.\ B {\bf 716} (2012) 30
  [arXiv:1207.7235 [hep-ex]].

\bibitem{Bardeen:1995kv}
  W.~A.~Bardeen,
  \emph{On naturalness in the standard model},
  FERMILAB-CONF-95-391-T.

\bibitem{Coleman:1973jx}
  S.~R.~Coleman and E.~J.~Weinberg,
  \emph{Radiative Corrections as the Origin of Spontaneous Symmetry Breaking},
  Phys.\ Rev.\ D {\bf 7} (1973) 1888.

\bibitem{Gildener:1976ih}
  E.~Gildener and S.~Weinberg,
  \emph{Symmetry Breaking and Scalar Bosons},
  Phys.\ Rev.\ D {\bf 13}, 3333 (1976).

\bibitem{Meissner:2006zh}
  K.~A.~Meissner and H.~Nicolai,
  \emph{Conformal Symmetry and the Standard Model},
  Phys.\ Lett.\ B {\bf 648} (2007) 312
  [hep-th/0612165].

\bibitem{Foot:2007as}
  R.~Foot, A.~Kobakhidze and R.~R.~Volkas,
  \emph{Electroweak Higgs as a pseudo-Goldstone boson of broken scale invariance},
  Phys.\ Lett.\ B {\bf 655} (2007) 156
  [arXiv:0704.1165 [hep-ph]].

\bibitem{AlexanderNunneley:2010nw}
  L.~Alexander-Nunneley and A.~Pilaftsis,
  \emph{The Minimal Scale Invariant Extension of the Standard Model},
  JHEP {\bf 1009}, 021 (2010)
  [arXiv:1006.5916 [hep-ph]].

\bibitem{Antipin:2013exa}
  O.~Antipin, M.~Mojaza and F.~Sannino,
  \emph{Natural Conformal Extensions of the Standard Model},
  arXiv:1310.0957 [hep-ph].

\bibitem{Espinosa:2007qk}
  J.~R.~Espinosa and M.~Quiros,
  \emph{Novel Effects in Electroweak Breaking from a Hidden Sector},
  Phys.\ Rev.\ D {\bf 76} (2007) 076004
  [hep-ph/0701145].

\bibitem{Hempfling:1996ht}
  R.~Hempfling,
  \emph{The Next-to-minimal Coleman-Weinberg model},
  Phys.\ Lett.\ B {\bf 379} (1996) 153
  [hep-ph/9604278].

\bibitem{Chang:2007ki}
  W.~-F.~Chang, J.~N.~Ng and J.~M.~S.~Wu,
  \emph{Shadow Higgs from a scale-invariant hidden U(1)(s) model},
  Phys.\ Rev.\ D {\bf 75} (2007) 115016
  [hep-ph/0701254 [HEP-PH]].

\bibitem{Iso:2009ss}
  S.~Iso, N.~Okada and Y.~Orikasa,
  \emph{Classically conformal $B-L$ extended Standard Model},
  Phys.\ Lett.\ B {\bf 676} (2009) 81
  [arXiv:0902.4050 [hep-ph]].

\bibitem{Englert:2013gz}
  C.~Englert, J.~Jaeckel, V.~V.~Khoze and M.~Spannowsky,
  \emph{Emergence of the Electroweak Scale through the Higgs Portal},
  JHEP {\bf 1304} (2013) 060
  [arXiv:1301.4224 [hep-ph]].

\bibitem{Hambye:2013dgv}
  T.~Hambye and A.~Strumia,
  \emph{Dynamical generation of the weak and Dark Matter scale},
  Phys.\ Rev.\ D {\bf 88} (2013) 055022
  [arXiv:1306.2329 [hep-ph]].

\bibitem{Carone:2013wla}
  C.~D.~Carone and R.~Ramos,
  \emph{Classical scale-invariance, the electroweak scale and vector dark matter},
  Phys.\ Rev.\ D {\bf 88} (2013) 055020
  [arXiv:1307.8428 [hep-ph]].

\bibitem{Foot:2007ay}
  R.~Foot, A.~Kobakhidze, K.~.L.~McDonald and R.~.R.~Volkas,
  \emph{Neutrino mass in radiatively-broken scale-invariant models},
  Phys.\ Rev.\ D {\bf 76} (2007) 075014
  [arXiv:0706.1829 [hep-ph]].

\bibitem{Foot:2007iy}
  R.~Foot, A.~Kobakhidze, K.~L.~McDonald and R.~R.~Volkas,
  \emph{A Solution to the hierarchy problem from an almost decoupled hidden sector within a classically scale invariant theory},
  Phys.\ Rev.\ D {\bf 77} (2008) 035006
  [arXiv:0709.2750 [hep-ph]].

\bibitem{Meissner:2008gj}
  K.~A.~Meissner and H.~Nicolai,
  \emph{Neutrinos, Axions and Conformal Symmetry},
  Eur.\ Phys.\ J.\ C {\bf 57} (2008) 493
  [arXiv:0803.2814 [hep-th]].

\bibitem{Holthausen:2009uc}
  M.~Holthausen, M.~Lindner and M.~A.~Schmidt,
  \emph{Radiative Symmetry Breaking of the Minimal Left-Right Symmetric Model},
  Phys.\ Rev.\ D {\bf 82} (2010) 055002
  [arXiv:0911.0710 [hep-ph]].

\bibitem{Dermisek:2013pta}
  R.~Dermisek, T.~H.~Jung and H.~D.~Kim,
  \emph{Coleman-Weinberg Higgs},
  arXiv:1308.0891 [hep-ph].

\bibitem{Hambye:2007vf}
  T.~Hambye and M.~H.~G.~Tytgat,
  \emph{Electroweak symmetry breaking induced by dark matter},
  Phys.\ Lett.\ B {\bf 659} (2008) 651
  [arXiv:0707.0633 [hep-ph]].



\bibitem{Foot:2010av}
  R.~Foot, A.~Kobakhidze and R.~R.~Volkas,
  \emph{Stable mass hierarchies and dark matter from hidden sectors in the scale-invariant standard model},
  Phys.\ Rev.\ D {\bf 82} (2010) 035005
  [arXiv:1006.0131 [hep-ph]].

\bibitem{Ishiwata:2011aa}
  K.~Ishiwata,
  \emph{Dark Matter in Classically Scale-Invariant Two Singlets Standard Model},
  Phys.\ Lett.\ B {\bf 710} (2012) 134
  [arXiv:1112.2696 [hep-ph]].

\bibitem{Farzinnia:2013pga}
  A.~Farzinnia, H.~-J.~He and J.~Ren,
  \emph{Natural Electroweak Symmetry Breaking from Scale Invariant Higgs Mechanism},
  Phys.\ Lett.\ B {\bf 727} (2013) 141
  [arXiv:1308.0295 [hep-ph]].

\bibitem{Gabrielli:2013hma}
  E.~Gabrielli, M.~Heikinheimo, K.~Kannike, A.~Racioppi, M.~Raidal and C.~Spethmann,
  \emph{Towards Completing the Standard Model: Vacuum Stability, EWSB and Dark Matter},
  Phys.\ Rev.\ D {\bf 89} (2014) 015017
  [arXiv:1309.6632 [hep-ph]].

\bibitem{Guo:2014bha}
  J.~Guo and Z.~Kang,
  \emph{Higgs Naturalness and Dark Matter Stability by Scale Invariance},
  arXiv:1401.5609 [hep-ph].

\bibitem{Hur:2011sv}
  T.~Hur and P.~Ko,
  \emph{Scale invariant extension of the standard model with strongly interacting hidden sector},
  Phys.\ Rev.\ Lett.\  {\bf 106} (2011) 141802
  [arXiv:1103.2571 [hep-ph]].

\bibitem{Heikinheimo:2013fta}
  M.~Heikinheimo, A.~Racioppi, M.~Raidal, C.~Spethmann and K.~Tuominen,
  \emph{Physical Naturalness and Dynamical Breaking of Classical Scale Invariance},
  arXiv:1304.7006 [hep-ph].

\bibitem{Holthausen:2013ota}
  M.~Holthausen, J.~Kubo, K.~S.~Lim and M.~Lindner,
  \emph{Electroweak and Conformal Symmetry Breaking by a Strongly Coupled Hidden Sector},
  JHEP {\bf 1312} (2013) 076
  [arXiv:1310.4423 [hep-ph]].

\bibitem{Ma:2013yga}
  E.~Ma, I.~Picek and B.~Radov\v{c}i\'c,
  \emph{New Scotogenic Model of Neutrino Mass with $U(1)_D$ Gauge Interaction},
  Phys.\ Lett.\ B {\bf 726} (2013) 744
  [arXiv:1308.5313 [hep-ph]].

\bibitem{Ade:2013zuv}
  P.~A.~R.~Ade {\it et al.}  [Planck Collaboration],
  \emph{Planck 2013 results. XVI. Cosmological parameters},
  arXiv:1303.5076 [astro-ph.CO].

\end{thebibliography}
\end{document}